\documentclass[a4paper,twoside]{article}

\newcommand{\affil}[1]{$^{\rm #1}$}

\usepackage[authoryear]{natbib}
\bibpunct{(}{)}{;}{a}{}{,}
\usepackage{graphicx}
\date{} 

\newcommand{\kms}{\mbox{km\,s$^{-1}$}}

\usepackage{amssymb}

\title{\large\bf\flushleft The Barium Isotopic Abundance in the Metal-Poor Star HD140283}
\author{\parbox{\textwidth}{\flushleft
\vspace{-0.5cm}
{\it R. Collet\affil{A,B}, M. Asplund\affil{A}, and P. E. Nissen\affil{C}} \\
\vspace{0.4cm}
{\small \affil{A}\, Max-Planck-Institut f{\"u}r Astrophysik, Postfach 1317, D--85741 Garching b. M{\"u}nchen, Germany}\\
{\small \affil{B}\,E-mail: remo@mpa-garching.mpg.de}\\
{\small \affil{C}\, Department of Physics and Astronomy, University of Aahrus, DK--8000 Aahrus C, Denmark}}}

\begin{document}

\twocolumn[
\begin{changemargin}{.8cm}{.5cm}
\begin{minipage}{.9\textwidth}
\vspace{-1cm}
\maketitle
%

\small{\bf Abstract:}
We derive the mixture of odd to even barium isotopes in the atmosphere of 
the metal-poor subgiant HD140283 from the analysis of the Ba~{\sc ii} transition 
at 4554~{\AA} in a high-resolution high signal-to-noise spectrum of the star.
The detailed shape of this spectral line depends on the relative contributions 
of odd and even isotopes via isotopic and hyperfine splitting. 
We measure the fractional abundance of odd Ba isotopes by modelling the formation of 
the Ba~{\sc ii}~4554~{\AA} line profile with the use of both a classical 1D hydrostatic
and a 3D hydrodynamical model atmosphere of HD140283.
We interpret the results in terms of contributions by the slow ($s$-) and rapid ($r$-)
neutron-capture processes to the isotopic mix.
While the result of the 1D analysis of the Ba~{\sc ii} feature indicates
a $64{\pm}36$\% contribution of the $r$-process to the isotopic mix,
the 3D analysis points toward a mere $15{\pm}34$\% contribution from this process,
that is consistent with a solar-like mixture of barium isotopes.


\medskip
\medskip
\end{minipage}
\end{changemargin}
]
\small

\section{Introduction}
Elements beyond the Fe peak are predominantly produced through successive 
neutron-capture reactions in two processes known as the slow ($s$-) and the 
rapid ($r$-) process.
The distinction between the $s$- and $r$- process depends on whether 
the time-scale for neutron captures is longer or shorter, respectively, 
than the time-scale of radioactive decay of freshly synthesized unstable nuclei.
Observational evidence and theoretical studies have identified the $s$-process 
site in low- to intermediate-mass (${\sim}1.3$--$8$~M$_\odot$) stars in the 
asymptotic giant branch (AGB).
In particular, the He-shell of thermally pulsating low-mass AGB stars is believed 
to be the site of the so-called \emph{main} $s$-process, which synthesizes nuclides 
heavier than Sr \citep[see, e.g.,][]{busso_gall_wasserb_1999}.
The $r$-process instead is usually associated with the explosive environment of
Type II supernovae (SNeII), although this astrophysical site hasn't been 
fully confirmed yet \citep[see, e.g.,][]{sneden_cowan_gallino_2008}.

Massive stars that end their evolution as SNeII are shorter-lived compared with 
low-mass stars that eventually evolve into the AGB phase.
One therefore expects the $r$-process to dominate the enrichment of the interstellar 
medium (ISM) in terms of heavy elements during the early stages of Galactic chemical 
evolution\footnote{Assuming, of course, that SNeII indeed are the
site of the $r$-process}.
According to this picture, the oldest most metal-poor stars should 
then only contain heavy elements in the relative proportions determined 
by the $r$-process \citep{truran_1981}.

\cite{magain_1995} used a novel approach to infer the relative contributions 
of the $r$- and $s$-process spectroscopically  by looking at the detailed shape 
of the Ba~{\sc ii}~$4554$~{\AA} resonance line.
The odd Ba isotopes contribute to broaden the line and alter the symmetry 
of its profile via hyperfine splitting. 
As the $r$- and $s$-process produce rather different mixtures of odd and even Ba 
isotopes, the actual width and shape of the  Ba~{\sc ii}~$4554$~{\AA} line are 
therefore dependent on the relative contribution of the two neutron-capture 
processes.
Based on the analysis of solar abundances \citep{anders89} by 
\cite{arlandini_1999}, one expects a value of $f_\mathrm{odd}^\mathrm{s}=0.11$ 
for the fractional abundance of the odd isotopes of barium\footnote{$f_\mathrm{odd}
{\equiv} [ N(^{135}\mathrm{Ba}) + N(^{137}\mathrm{Ba}) ] / N(\mathrm{Ba})$},
in case of a pure $s$-process mixture, and $f_\mathrm{odd}^\mathrm{r}=0.46$ in case
of a pure $r$-process one.
Magain challenged the Galactic chemical evolution scenario depicted above by 
deriving a fractional abundance of odd Ba isotopes $f_{\mathrm{odd}}=0.06{\pm}0.06$ 
for the metal-poor subgiant HD140283 ([Fe/H]$=-2.5$): this value is fully compatible
with a pure $s$-process production and excludes any significant contribution from 
the $r$-process.
Magain's result cannot indeed be reconciled with the predictions of standard 
Galactic chemical evolution models \citep[e.g.,][]{travaglio_1999} which indicate 
that the $s$-process signature in stars should become manifest only at higher 
metallicities ([Fe/H]${\gtrsim}-1.5$).
More recently, \cite{lap_2002} re-observed and re-analysed the Ba~{\sc ii}~$4554$~{\AA} 
line in the spectrum of HD140283 and derived, in contrast to Magain, a fractional 
abundance of odd Ba isotopes $f_{\mathrm{odd}}=0.30{\pm}0.21$, consistent with a 
pure $r$-process isotopic mixture.
\cite{lap_2002} used a spectrum of superior quality than Magain's both 
in terms of resolution (${\lambda}/{\Delta}{\lambda}{\approx}200{\,}000$) and 
signal-to-noise ratio (S/N${\approx}550$); their analysis, however, was still 
based on 1D model stellar atmospheres.

Classical spectroscopic analyses based on 1D model atmospheres, however, 
rely on a number of tunable fudge parameters. In particular, they
cannot account for Doppler broadening due to photospheric convective motions
without introducing the micro- and macro-turbulence parameters.
Moreover, convective flows also induce asymmetries and overall wavelength shifts 
in spectral line profiles, which cannot be reproduced by ordinary 1D analyses.

Three-dimensional hydrodynamical simulations of stellar surface convection, 
on the other hand, can self-consistently predict photospheric velocity fields 
and correlated temperature and density inhomogeneities \citep{stein98,asplund99};
line shapes, asymmetries, and wavelength shifts can be accurately reproduced 
by using such simulations as 3D hydrodynamical model atmospheres, without relying 
on ad hoc free parameters \citep{asplund00fe2}.
In the present contribution, we re-derive the fractional abundance of the odd 
Ba isotopes in HD140283 by means of both a 1D and a 3D analysis of the 
Ba~{\sc ii}~$4554$~{\AA} line in the spectrum obtained by \cite{lap_2002} at the 
W.~J.~McDonald Observatory (Mt Locke, Texas).
In particular, we use a 3D model atmosphere of the metal-poor subgiant star to 
adequately disentangle the contributions of photospheric convective flows and 
hyperfine splitting to the broadening and asymmetry of the Ba~{\sc ii}~$4554$~{\AA} 
line.

\begin{table}[ht]
\begin{center}
\caption{Isotopic and hyperfine splitting components of the Ba~{\sc ii} line at 4554~{\AA}.
The components have been computed using the atomic data by \cite{wendt84} and \cite{villemoes93}
for the splitting of the lower and upper energy levels.
The oscillator strengths relative to the $^{138}$Ba component are given in the third column.
The adopted ${\log}{gf}$ value for the line is $+0.17$.}
\label{tab:hfscomp}
\begin{tabular}{lcc}
\smallskip \\
\hline 
Isotope & Wavelength ({\AA}) & Relative strength\\
\hline
$^{134}$Ba: 	&	4554.0314 &	1.0000 \\
\medskip \\
$^{135}$Ba:
		&	4554.0003 &    0.1563 \\
       		&	4554.0015 &    0.1562 \\
	       	&	4554.0019 &    0.0625 \\
       		&	4554.0473 &    0.4375 \\
       		&	4554.0500 &    0.1563 \\
       		&	4554.0512 &    0.0313 \\
\medskip \\
$^{136}$Ba:	&	4554.0317 &    1.0000 \\       
\medskip \\
$^{137}$Ba:
     		&       4553.9975 &    0.1563 \\
       		&       4553.9986 &    0.1562 \\
       		&       4553.9988 &    0.0625 \\
			&		4554.0498 &    0.4375 \\
       		&		4554.0531 &    0.1563 \\
       		&       4554.0542 &    0.0313 \\
\medskip \\
$^{138}$Ba:	&       4554.0330 &    1.0000 \\
\hline
\end{tabular}
\end{center}
\end{table}

\section{Methods}
\label{sec:methods}
We synthesize the Ba~{\sc ii}~$4554$~{\AA} line profile with the help of 
both a 1D and a 3D model atmosphere of HD140283, varying the Ba isotopic 
mixture to produce the best fit to the observed feature.
We account for the isotopic and hyperfine splitting components of the 
Ba~{\sc ii}~$4554$~{\AA} line listed in table~\ref{tab:hfscomp}.
Isotopic wavelength shifts are actually too small to be resolved even with
the highest-resolution spectrographs currently available: in the present 
analysis, therefore, we effectively do not make any distinction between the
even isotopes as well as between the three odd isotopes.

For the 1D analysis, we employ a plane-parallel hydrostatic LTE {\sc marcs} 
model atmosphere \citep{gustafsson75,asplund97} of the subgiant star with 
the following stellar parameters: $T_\mathrm{eff}=5690$~K, ${\log}{g}=3.67$~(cgs),
and [Fe/H]$=-2.50$. 
Spectral line profiles are computed under the assumption of LTE.
We stress that, contrary to the overall barium abundance, 
isotopic abundances derived from the Ba~{\sc ii} $4554$~{\AA} line 
are expected to be insensitive to departures from LTE and to  
the actual choice of stellar parameters.
In the 1D calculations, we adopt a micro-turbulence $\xi=1.49$~{\kms}, 
based on the LTE analysis of the $51$~Fe~{\sc i} and $13$~Fe~{\sc ii} 
lines given in Tab.~2 of \cite{lap_2002}.
With the above value, the iron abundance becomes independent from the
equivalent width and chemical equilibrium is fulfilled
(${\log}{\epsilon}(\mathrm{Fe~{I}}) =4.98{\pm}0.13$ and 
 ${\log}{\epsilon}(\mathrm{Fe~{II}})=4.98{\pm}0.11$).

Before proceeding with the synthesis of the Ba~{\sc ii}~$4554$~{\AA} line we 
need to estimate the broadening from mechanisms other than hyperfine splitting.
In order to do this, we fit the Fe lines from the same sample used by 
\cite{lap_2002} (Tab.~3 in their article) by assuming a rotational velocity
$v_\mathrm{rot}{\sin}{i}=0.5$~{\kms} and varying the iron abundance, the central wavelength,
and the FWHM of a Gaussian which we use toconvolve the line profiles.
The convolution with a Gaussian accounts for macro-turbulent and 
instrumental broadening (and, possibly, residual rotational broadening).
We account for natural and linear Stark broadening and we model collisional 
broadening with neutral hydrogen atoms using the quantum mechanical calculations 
by \cite{barklem00}.
From the analysis of the Fe lines, we derive an average value of 
$4.87{\pm}0.11$~{\kms} for the FWHM of the Gaussian broadening.
Using this estimate of the Gaussian broadening, we compute synthetic 
flux profiles of the Ba~{\sc ii}~$4554$~{\AA} line for different barium 
isotopic mixtures.
We quantify the comparison between the theoretical and observed profiles
by means of a $\chi^2$-analysis similar to the one carried out by
\cite{asplund99} to investigate $^6$Li/$^7$Li ratios in metal-poor halo stars.
We compute the $\chi^2$ according to the expression 
$\chi^2={\sum}(O_\mathrm{i}-S_\mathrm{i})^2/\sigma^2$, with $O_\mathrm{i}$
and $S_\mathrm{i}$ denoting the observed and synthetic flux at
wavelength point $i$, respectively, and $\sigma$ is the inverse
signal-to-noise ratio.
The most likely isotopic mix is the one that minimizes $\chi^2$.

\begin{figure}[ht]
\begin{center}
\includegraphics[scale=0.45, angle=0]{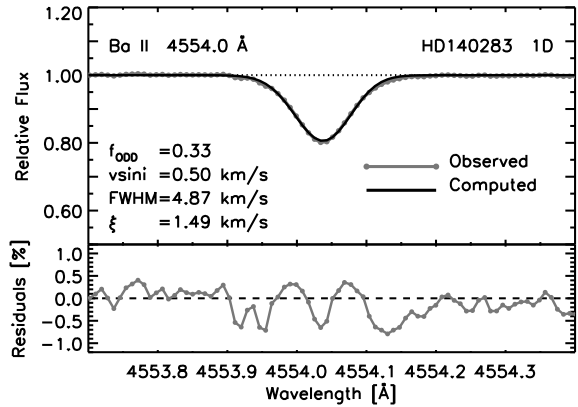}
\caption{\emph{Upper panel}: Synthetic (\emph{continuous black line})
versus observed (\emph{grey symbols}) profiles of the Ba~{\sc ii} line at
$4554.0$~{\AA} for the best fitting Ba isotopic mix in the 1D analysis.
\emph{Lower panel}: Relative difference between observed and synthetic profiles.}
\label{fig:bestfit_1d}
\end{center}
\end{figure}
\begin{figure}[ht]
\begin{center}
\includegraphics[scale=0.49, angle=0]{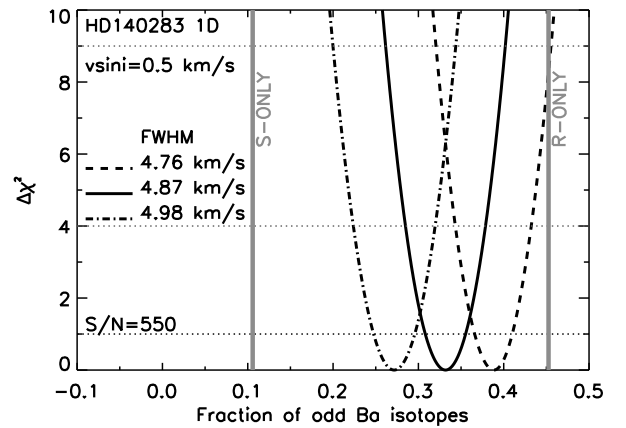}
\caption{\emph{Thick continuous line}: 
Resulting $\Delta\chi^2=\chi^2-\chi_{\mathrm{min}}^2$ from the fit 
of the Ba~{\sc ii}~$4554$~{\AA} line as a function of the fractional 
abundance of odd Ba isotopes for the combined Gaussian (FWHM$=4.87$~{\kms}) 
of the mean macro-turbulence and instrumental broadening in the 1D case;
also shown are the cases for FWHM${\pm}{\Delta}{\mathrm{FWHM}}$
(\emph{thick dashed} and \emph{dot-dashed lines}). 
A value $v_{\mathrm{rot}}{\sin}{i}=0.5$~{\kms}~ is adopted for the 
rotational velocity.
\emph{Thick grey lines}: fractional abundances of odd Ba isotopes 
in case of pure $s$-process and $r$-process isotopic mixes 
(S.~Bisterzo 2008, private communication). }
\label{fig:chi2_1d}
\end{center}
\end{figure}

For the 3D analysis, we use a simulation of  stellar surface convection 
by \cite{asplund99}, constructed for the same stellar parameters as for the 
1D case, as a time-dependent 3D hydrodynamical model atmosphere.
We select a $35$-minute long sequence of $30$~simulation snapshots sampled
at regular time intervals and downgrade the resolution of the original 
simulation to $50{\times}50{\times}82$ grid-points prior to the spectral 
line formation calculations.
We compute LTE flux profiles in 3D by solving the radiative transfer equation
along $33$ rays (four $\mu$-angles, eight $\phi$-angles, and the vertical) 
for all grid-points at the surface, performing then a disk integration and 
a time average over all snapshots. 
As we already account for the effect of Doppler shifts induced by the velocity 
fields in the 3D simulation, we ignore altogether the micro- and 
macro-turbulence parameters.
We then determine the amount of broadening contributed 
by processes other than hyperfine splitting: we fit the profiles of 
the Fe lines similarly as in the 1D analysis above but varying the 
rotational velocity instead of the Gaussian broadening, which, in 3D,
we keep fixed and equal to the instrumental broadening (corresponding 
roughly to a FWHM$=1.5$~{\kms}, as estimated from the resolution of the 
observed spectrum).
Natural and collisional broadening are included and treated in the same way 
as in the 1D calculations.
From the 3D analysis of the Fe lines, we estimate an average rotational velocity
of $2.58{\pm}0.30$~{\kms}.
We then follow the same $\chi^2$-minimization procedure as in the 1D analysis
to determine the most probable barium isotopic mix from the fitting of
the observed Ba~{\sc ii}~$4554$~{\AA} line profile.

\section{Results}
\subsection{1D analysis}

Figure~\ref{fig:bestfit_1d} shows the best-fitting synthetic profile
to the observed Ba~{\sc ii} $4554.0$~{\AA} line in the 1D case. 
The derived fractional abundance of odd Ba isotopes in 1D is 
$f_\mathrm{odd}=0.33$, which suggests a contribution predominantly 
from the $r$-process (namely $64$\%) to the barium isotopic mix.
Figure~\ref{fig:chi2_1d} shows the $\Delta\chi^2=\chi^2-\chi_{\mathrm{min}}^2$
curves considered in the $\chi^2$-minimization procedure.
The $1\sigma$, $2\sigma$, and $3\sigma$ confidence limits correspond to 
$\Delta\chi^2=1$, $4$, and $9$, respectively (indicated by the dotted lines
in the figure).
The half-width of the $f_\mathrm{odd}$ interval for which $\Delta\chi^2<1$ 
is a measure of the uncertainty in fractional abundance of odd isotopes
due to the finite signal-to-noise; for S/N$=550$ the uncertainty on 
$f\mathrm{odd}$ is ${\pm}0.023$.
Figure~\ref{fig:chi2_1d} also shows how the position of the $\chi^2$ 
minimum depends on the Gaussian broadening: a change of ${\pm}0.11$~{\kms} 
translates into a change of ${\mp}0.06$ in $f_\mathrm{odd}$.
The fractional abundance of odd Ba isotopes is also sensitive to the choice
of micro-turbulence parameter (which affects the Gaussian broadening
determination) and to the adopted surface gravity (which affects pressure 
broadening of the lines). 
For the present analysis, we adopt the estimates of ${\mp}0.11$ and ${\mp}0.01$
by \cite{lap_2002} of the errors on $f_\mathrm{odd}$ due to the uncertainties on 
micro-turbulence (${\pm}0.2$~{\kms}) and ${\log}{g}$ (${\pm}0.2$~{dex}), 
respectively.
Our estimated total uncertainty on $f_\mathrm{odd}$ is then
${\Delta}f_\mathrm{odd}=${\linebreak[4]}$\sqrt{0.023^2+0.06^2+0.11^2+0.01^2}$ 
${\approx}0.13$ which directly corresponds to an uncertainty in the contribution of 
$r$-process fraction of Ba isotopes of about ${\pm}36$\%.

\subsection{3D analysis}

\begin{figure}[ht]
\begin{center}
\includegraphics[scale=0.45, angle=0]{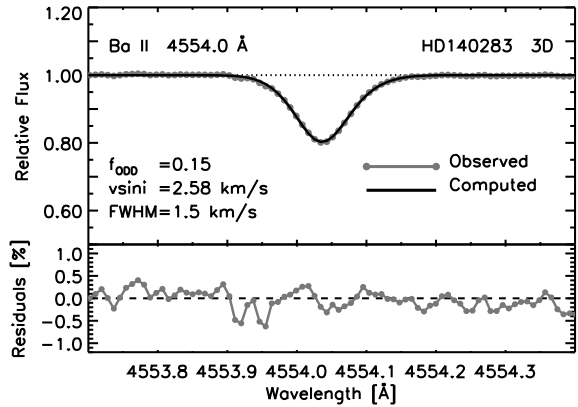}
\caption{Same as Fig.~\ref{fig:bestfit_3d} but for the best fitting Ba 
isotopic mix in the 3D analysis.}
\label{fig:bestfit_3d}
\end{center}
\end{figure}

\begin{figure}[ht]
\begin{center}
\includegraphics[scale=0.49, angle=0]{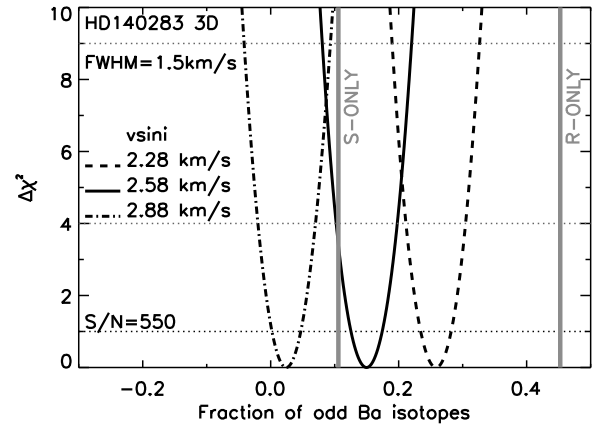}
\caption{\emph{Thick continuous line}:
Resulting $\Delta\chi^2$ from the fit of the Ba~{\sc ii}~$4554$~{\AA}
line as a function of the fractional abundance of odd Ba isotopes for a 
rotational velocity ${v_\mathrm{rot}\sin{i}}=2.58$~{\kms} in the 3D case;
also shown are the cases for ${v_\mathrm{rot}\sin{i}}{\pm}{\Delta}{v_\mathrm{rot}\sin{i}}$
(\emph{thick dashed} and \emph{dot-dashed lines}). 
A value FWHM$=1.5$~{\kms}~ is assumed for the instrumental broadening.}
\label{fig:chi2_3d}
\end{center}
\end{figure}

The best-fitting synthetic Ba~{\sc ii} $4554.0$~{\AA} line profile
in the 3D analysis is shown in Fig.~\ref{fig:bestfit_3d}. 
It is clear from the comparison of the residuals, that the synthetic
profile in 3D gives a significantly better fit to the observed spectral line
than the synthetic 1D profile.
The estimated fractional abundance of odd Ba isotopes in the 3D analysis is
$f_\mathrm{odd}=0.15$, which ---contrary to the 1D case--- points toward
a predominantly $s$-process isotopic mix (only $12$\% contribution from the
$r$-process).
Figure~\ref{fig:chi2_3d} shows the $\Delta\chi^2=\chi^2-\chi_{\mathrm{min}}^2$
curves considered for the $\chi^2$-minimization procedure in the 3D analysis.
Similarly as for the 1D case, we use the information in the figure
to quantify the uncertainties on $f_\mathrm{odd}$ due to the finite signal-to-noise
and the error on the estimated rotational velocity: these are $0.024$
and $0.12$, respectively.
For the error on $f_\mathrm{odd}$ due to the uncertainty on surface gravity, 
we adopt the same value as in the 1D case. 
We caution that a change in surface gravity in practice also affects the
velocities predicted by the simulation. In this preliminary 3D analysis of the
Ba~{\sc ii} $4554.0$~{\AA} line, however, we neglect this aspect;
we defer the study of the effects of changes in stellar parameters on the
estimate of the uncertainty on $f_\mathrm{odd}$ to a future and more comprehensive work.
Finally, we do not consider any contribution to the error on $f_\mathrm{odd}$
from micro-turbulence, since this parameter is absent in the 3D analysis.
 The estimated total uncertainty on $f_\mathrm{odd}$ in the 3D analysis is then
${\Delta}f_\mathrm{odd}=$ $\sqrt{0.024^2+0.12^2+0.01^2}$ ${\approx}0.12$
which directly corresponds to an uncertainty in the contribution of 
$r$-process fraction of about ${\pm}34$\%.

\section{Discussion}
The fractional abundance $f_\mathrm{odd}=0.33{\pm}0.13$ of odd Ba
isotopes derived in the present 1D analysis is in excellent
agreement with the result obtained by {\linebreak[4]}\cite{lap_2002}.
Contrary to the finding of \cite{magain_1995}, our 1D estimate of 
$f_\mathrm{odd}$ indicates a large contribution from the $r$-process
($64$\%) to the barium isotopic mix in HD140283 and seems to exclude
at a 3$\sigma$ level a pure $s$-process contribution.

Interestingly, the 3D analysis of the same feature leads to a radically
different result, namely that the Ba~{\sc ii} $4554.0$~{\AA} line
profile is best-fitted assuming instead a solar-like barium isotopic mix 
with only a $15$\% (${\pm}34$\%) contribution from the $r$-process.
Moreover, in 3D, a pure $r$-process isotopic mixture can be excluded
at a 2$\sigma$ level from the $\chi^2$ analysis.
The reason for the difference between the results of the 1D and 3D analyses
is essentially due to the differences between the 1D and 3D modelling of
line broadening and asymmetries as sketched in Sec.~\ref{sec:methods}.
For instance, in the 3D calculations, the simulation's velocity fields 
naturally induce a ``C-shaped'' asymmetry in the flux profile of the line.
This implies that a different fraction of odd barium isotopes 
is necessary in 3D comparing to the 1D case to properly model the shape
of the Ba~{\sc ii} $4554.0$~{\AA} line profile.
We would like to caution at this point the reader that the our result
for HD140283 cannot be applied straightforwardly to the 1D analysis of the Ba~{\sc ii} $4554.0$~{\AA} feature in the spectra of other stars. 
The sign and magnitude of the 3D$-$1D correction to the derived fractional 
abundance of odd isotopes may depend in general on the strength 
and detailed shape of the observed Ba~{\sc ii} $4554.0$~{\AA} profile.

Our 3D analysis of the Ba~{\sc ii} $4554.0$~{\AA} feature in the spectrum
of HD140283 seems unable to settle down the controversy raised by \cite{magain_1995}
in favour of standard Galactic chemical evolution scenarios.
On the contrary, our result seems to corroborate the possibility that
the metal-poor star HD140283 possesses a strong $s$-process signature 
in terms of isotopic Ba abundance against the expectations from theoretical models
of Galactic chemical evolution.
We would like to stress however that, although our and Magain's result are
apparently in agreement with each other, our finding is based on rather different
premises. 
First, we rely on a higher-resolution and higher signal-to-noise spectrum,
Second, we use a 3D hydrodynamical model atmosphere of the star to synthesize
the profiles of the Ba~{\sc ii} $4554.0$~{\AA} feature and other Fe lines,
which implies that our modelling of non-thermal broadening is more robust
and, contrary to the 1D analysis, independent on the tunable micro- and macro-turbulence parameters.

As a final note, we cannot completely rule out the possibility that HD140283 
has been polluted by $s$-process material from an AGB star and 
may not be representative of halo stars at the same metallicity in terms
of isotopic Ba abundance. 
Further investigation is therefore necessary to draw any significant comparison with
Galactic chemical evolution models.
We intend to extend in an upcoming paper the 3D$-$1D analysis of the 
Ba~{\sc ii} $4554.0$~{\AA} feature to a larger sample of halo stars
and study in particular the barium isotopic abundance as a function of metallicity.

\section*{Acknowledgments} 
The authors would like to thank C. Allende Prieto for
kindly making parts of the McDonald spectrum of HD140283 available to us
for the present analysis.

\bibliographystyle{apj}
\bibliography{collet}{}


\end{document}